\documentclass[preprint,aps]{revtex4-1}
\usepackage{graphicx}
\usepackage{bm}

\newcommand{\cout}[1] { }
\def\vzlle{ \langle 2\bar{1}\bar{1}3\rangle /3}

\def\polll{ \{ 01\bar{1}1 \} }
\def\pzllz{ \{ 2\bar{1}\bar{1}2 \} }

\def\plll{ \{ 111 \} }
\def\ploo{ \{ 100 \} }
\def\va{ \langle a \rangle }
\def\vc{ \langle c \rangle }
\def\vca{ \langle c+a \rangle }
\def\ab{ {\it ab-initio }}

\def\pl {PI }
\def\pz {PII }

\begin{document}

\title{
Novel Cross-Slip Mechanism of Pyramidal Screw Dislocations in Magnesium
}
\author{Mitsuhiro Itakura}
\author{Hideo Kaburaki}
\author{Masatake Yamaguchi}
\author{Tomohito Tsuru}
\affiliation{
Center for Computational Science \& e-Systems, Japan Atomic Energy Agency.
178-4-4 Wakashiba, Kashiwa, Chiba 277-0871, Japan }
\affiliation{
Center for Computational Science \& e-Systems, Japan Atomic Energy Agency.
2-4 Shirakata-Shirane, Tokai-mura, Naka-gun, Ibaraki 319-1195, Japan}
\affiliation{
 Nuclear Science and Engineering Center, Japan Atomic Energy Agency.
2-4 Shirakata-Shirane, Tokai-mura, Naka-gun, Ibaraki 319-1195, Japan
}

\begin{abstract}
Compared to  cubic metals, whose primary slip mode
includes twelve equivalent systems,
the lower crystalline symmetry of hexagonal close-packed metals
results in a reduced number of equivalent primary slips
and anisotropy in plasticity, leading to brittleness at the ambient temperature.
At higher temperatures, the ductility of hexagonal close-packed metals
improves owing to
the activation of secondary $\vca$ pyramidal slip systems.
Thus understanding the fundamental
properties of corresponding dislocations is essential for the improvement of ductility
at the ambient temperature.
Here, we present the results of large-scale \ab
calculations for $\vca$ pyramidal screw dislocations in Mg and show that
their slip behavior is a stark counterexample to the conventional wisdom that 
a slip plane is determined by the stacking fault plane of dislocations.
A stacking fault between dissociated partial dislocations
can assume a non-planar shape with a negligible energy cost
and can migrate normal to its plane by a local shuffling of atoms.
Partial dislocations dissociated on a $\pzllz$ plane
``slither'' in the $\polll$ plane, dragging the stacking fault with them 
in response to an applied shear stress.
This finding resolves the apparent discrepancy that both $\pzllz$ and $\polll$ slip traces
are observed in experiments while \ab calculations indicate that dislocations preferably dissociate in the $\pzllz$ planes.

\end{abstract}
\pacs{Valid PACS appear here}
\maketitle


\newcommand{\figl}[1] { {\bf (#1)} }


The deformation of crystalline materials is induced by the glide of dislocations and migration of twin boundaries, and the anisotropy
of deformation is determined by the atomistic structures of these defects \cite{textbook}.
Compared to  cubic metals, the plastic anisotropy of hexagonal metals, such as Magnesium(Mg), zirconium, and titanium,
is much more pronounced owing to their lower crystalline symmetry and fewer easy slip systems.
At present, press-forming of hexagonal metals is only possible at elevated temperatures, where 
the plasticity becomes less anisotropic owing to the activation of secondary slip systems.
This limits the use of hexagonal metals with excellent specific strength such as Mg \cite{mg-review}
as structural materials.
In order to improve their ductility at ambient temperature,
we must first identify the fundamental properties of dislocations associated with secondary slip systems.

In Mg and Mg-based alloys, the secondary slip system responsible for  improved ductility
has been identified
as the slip of $\vca$ pyramidal dislocations by observations of strained samples
\cite{obara73,mathis04,agnew05,sandlobes11,sandlobes12}.
The slip plane and the core structure of these dislocations, however, remain controversial issues.
The slip plane is experimentally identified by the slip trace angles.
An early study identified the traces as $\pzllz$ pyramidal II (\pz) plane \cite{obara73};
however, the result was recently reanalyzed and the traces were identified as $\polll$ pyramidal I (\pl)
plane \cite{fan15}.
Both \pl and \pz traces were observed in an MgY alloy \cite{sandlobes11,sandlobes12}.
Recent observation of pure Mg showed that the dominant slip traces are \pl
and the slip trace angles vary continuously, indicating frequent cross-slips \cite{xie16}.
Recent atomistic calculations report that
$\vca$ dislocations can dissociate on both \pl and \pz planes \cite{nogaret10,ghazi-dft,tang14}, whereas
dissociation is more stable on the \pz plane than on the \pl plane \cite{ghazi-dft,wu15md}.
The atomistic calculations of \pz edge dislocation
explain the peculiar hardening behavior in Mg \cite{wu15md2}.
However, slips of $\vca$ dislocations on both \pl and \pz planes are reported in 
molecular dynamics (MD) simulations,
and cross-slips are also observed at moderate temperatures \cite{tang14,tang14-2}.

Thus, the cross-slip behavior of the $\vca$ dislocations in Mg is rather singular,
from the conventional perspective summarized as follows:
(i) dislocations dissociate into partials on a particular plane
and their motion is confined to that plane; (ii) a change of the slip plane
is attributed to a thermally activated cross-slip, which involves the recombination of partials in the screw orientation;
(iii) frequent cross-slip is only possible at very high temperatures.
In the present work, we closely investigate the slip behavior of
$\vca$ pyramidal screw dislocations using \ab calculations
and show that the origin of the singular cross-slip behavior
is the migration of partials in the direction
normal to the plane of dissociation.

\begin{figure}[h]

\includegraphics[width=17cm, bb=0 0 1136 358]{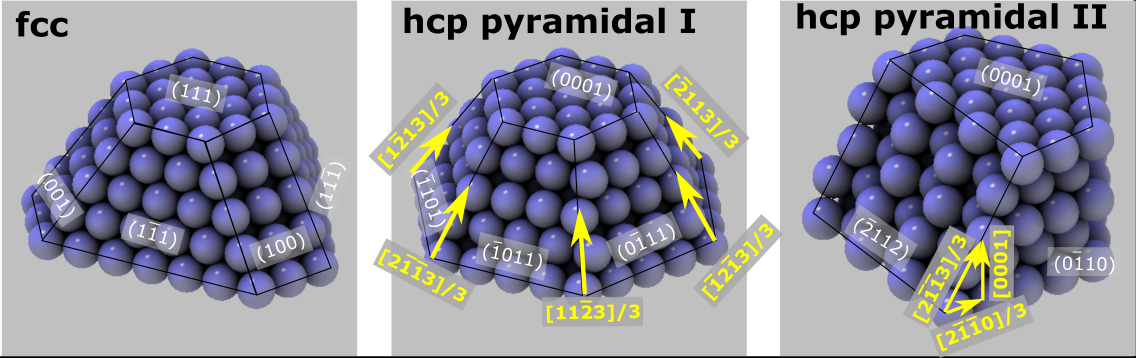}
\caption{
 The crystal structures and slip planes of fcc and hcp crystals, constructed from close-packed spheres.  Burgers vectors and slip planes are indicated by yellow and white letters, respectively}
\label{fign-fcchcp}
\end{figure}

Figure \ref{fign-fcchcp}
shows close-packed slip planes in face-centered cubic (fcc) metals and the corresponding slip planes in hexagonal-close packed (hcp) metals with the ideal
$c/a$ ratio.
While there are four possible close-packed $\plll$ slip planes in the fcc structure, the hcp structure  has only one close-packed plane, i.e., the basal plane $(0001)$. The six equivalent $\polll$ \pl planes comprise  alternating stacks of close-packed $\plll _{fcc}$-like units and non-close-packed $\ploo _{fcc}$-like units. The $\pzllz$ \pz plane
is more jagged compared to \pl because 
it comprises alternating \pl planes on the atomistic scale.
The Burgers vector on these planes, $\vca = \vzlle$, is approximately twice as long as $\va$. The activation of slip systems with a Burgers vector of $\vca$ requires much higher temperatures compared to those with a Burgers vector of $\va$. 
Because any slip associated with $\va$ dislocations has a zero Schmid factor for loading in the $\vc$ direction,
slips of  $\vca$ dislocations are indispensable in the deformation of a polycrystal, wherein  some of the grains 
align their $\vc$ axis to the loading direction.
The  $\vca$ dislocation activity and twinning are usually observed in 
Mg-based alloys under $\vc$ axis loading \cite{obara73,barnet07}.

The  properties of dislocation are closely related to the generalized stacking fault energy (GSFE) 
of the slip plane.
We have calculated GSFE of \pl and \pz planes using \ab method and found that
there are metastable points in both cases,
whose approximate positions are $0.40 \langle 0\bar{1}12 \rangle /2$ and $0.48 \vca$ for  \pl and \pz , respectively \cite{note:twin-vector}.
The corresponding GSFE is $181 mJ/m^2$ and $207 mJ/m^2$, respectively \cite{note:gsfe}.
The existence of stable stacking fault (SF) on both slip planes indicates
that the dislocation dissociate into two partials on both planes. The partials are separated by a strip of the SF with a finite width $d$.
The dislocation line energy owing to the dissociation is, within an approximation of the isotropic linear elasticity, 
    given by
\begin{equation}
  -\frac{\mu B_2 }{2\pi} \log (d/r_c) +d\gamma_{SF},
  \label{eq:width}
\end{equation} 
where $\mu$ is the shear modulus,
$B_2 = b_1^s b_2^s + b_1^e b_2^e/(1-\nu)$,
$b_p^s$ and $b_p^e$ are screw and edge components of
Burgers vector of partial $p$, $\nu=0.35$ is the Poisson's ratio of Mg,
$r_c$ is core radius,
and $\gamma_{SF}$ is the stacking fault energy \cite{textbook}.
Minimizing  Eq. (\ref{eq:width}) with respect to $d$ gives  $d = \mu B_2 /2\pi \gamma_{SF}$.
Based on the anisotropic elastic constants reported in \cite{elas},
we use a shear modulus $\mu=20$ GPa \cite{note:aniso}.
With the calculated values of  $\gamma_{SF} = 0.18 J/m^2$ and $0.21 J/m^2$,  
 $B_2=8.07 $ and $9.26$ \AA$^2$
for \pl and \pz, respectively, we obtained $d\sim 14$ \AA
for both \pl and \pz cases.

\begin{figure}
\includegraphics[width=10cm, bb=0 0 1050 1043]{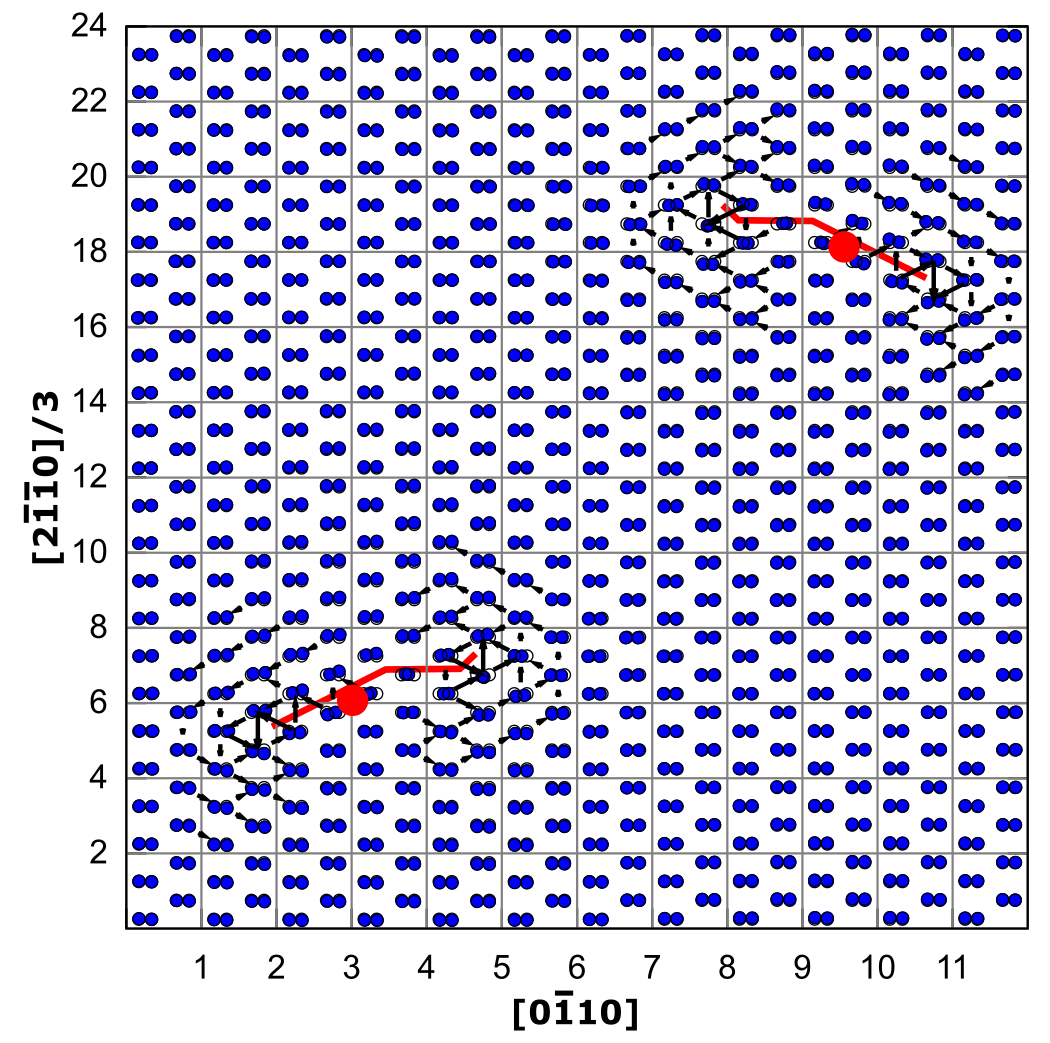}
\caption
{ Atomistic configuration of the periodic dislocation array used for the \ab calculations, seen 
    from the $\vca$ direction.
	Empty and filled circles are positions of atoms for the perfect
	crystal and the dislocation configuration, respectively. Bold lines show
	the stacking faults, and arrows show the differentiated displacement.
	Two red circles mark the positions of two dislocation cores of opposite sign.
} 
\label{fign-cel}
\end{figure}

The \ab calculations are performed with the
Vienna Ab-initio Simulation Package \cite{vasp1,vasp2} with the projector augmented wave method. The exchange correlation energy is calculated by the generalized gradient approximation with the Perdew-Burke-Ernzerhof function \cite{perdew96}. The Methfessel-Paxton smearing method with a width of 0.2-eV is also used. The cutoff energy for the plane-wave basis set is 210 eV.  Structural relaxation is terminated when the maximum force becomes less than $1$ meV/\AA.
Periodic quadrupolar array configuration of the dislocation core \cite{cai03}
is used to calculate core structures and their energies.
We use a skewed calculation cell
$L [0\bar{1}10] \times 2L [2\bar{1}\bar{1}0]/3  \times [2\bar{1}\bar{1}3]/3$ 
with $L=12$ which contains $1152$ atoms, as shown in Fig. \ref{fign-cel}. 
The lattice constants are $a_0=3.1935$ \AA
and $c_0=5.1776$ \AA.
The cell is sheared to cancel out the strain induced by the dipole moment of two cores \cite{cai03}. 
A $1\times 1 \times 4$ Monkhorst Pack k-point mesh is used.
The overall numerical uncertainty in the dislocation line energy arising from
the finite k-point mesh and the cutoff energy is $1.5$ meV/\AA.
The initial configurations are prepared using isotropic linear elasticity solutions.

The difference in the total energy between two types of
core structure in a finite cell
differ from that of isolated dislocations
owing to elastic interactions between
the mirror images, and a smaller cell size
results in greater modification.
The overall effect can be calculated by
an infinite summation of the mirror interactions between
partials \cite{note:converge}.
The overall finite size effect is estimated as
$0.5$ meV/\AA for the $L=12$ case.
Together with the
numerical uncertainty, the overall accuracy of the relative dislocation line energy
is $2$ meV/\AA.

\begin{figure}
\includegraphics[width=12cm, bb=0 0 787 1164]{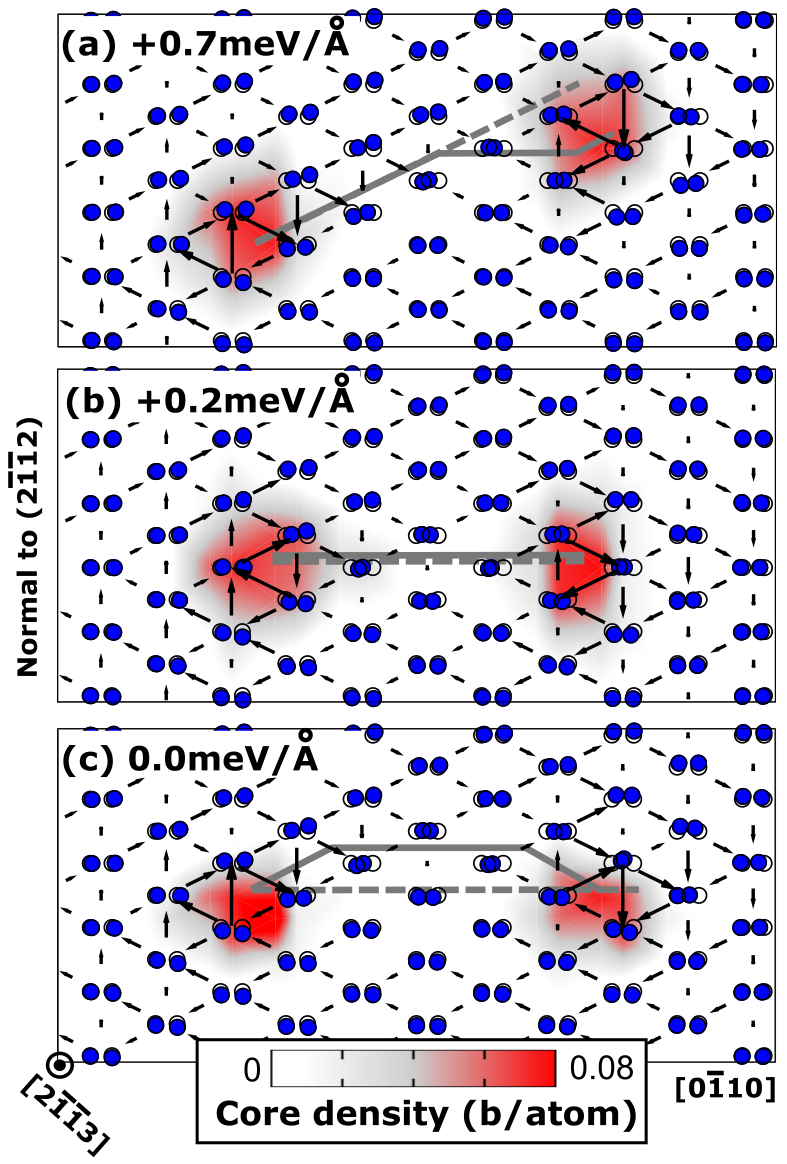}
\caption{ 
Core structures of $\vca$ pyramidal screw dislocations obtained by the \ab calculations.
Arrows show differentiated displacement in the $\vca$ direction by their length. Dashed and solid lines mark the position of SF before and after relaxation. The contour map shows the distribution of core density per atom, evaluated from the screw component of the Nye tensor and normalized so that
the summation over atoms gives unity. The relative line energies of each structure are also shown.
}
\label{fign-ddmap} 
\end{figure}

Figure \ref{fign-ddmap} shows stable and metastable core structures of
$\vca$ screw dislocation obtained by the \ab calculations using various initial configurations
with different dissociation planes and widths.
The core density distribution is evaluated by
the Nye tensor \cite{nye}, using the Atomsk software \cite{atomsk}.
The width of the SF is approximately $14$ to $15$ \AA, which
is consistent with Eq.(\ref{eq:width}).
In configurations (a) and (c),
a part of the SF migrates during the relaxation, resulting in non-planar shapes \cite{note:sf}.
The configuration (c) has the lowest energy, and its structure closely resembles the one reported in \cite{ghazi-dft}.
The line energy differences between three configurations are within the numerical accuracy of $2$ meV/\AA.

\begin{figure}[h]
\includegraphics[width=10cm, bb=0 0 1044 804]{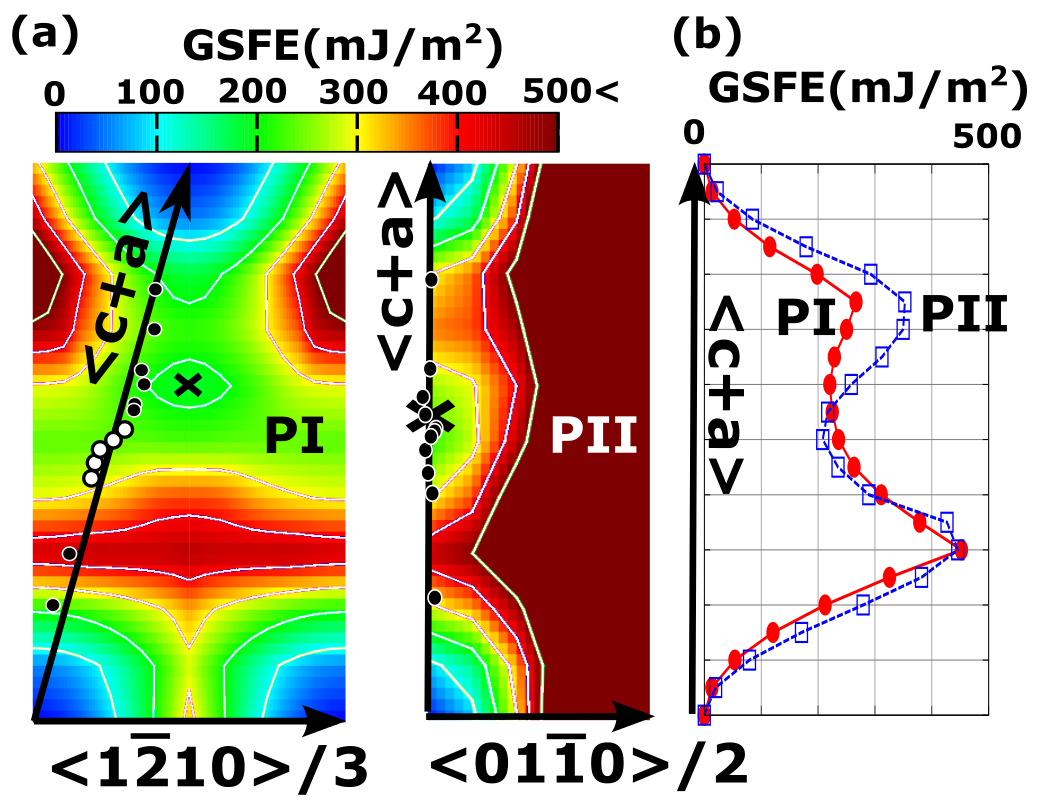}
\caption{
 (a) The distributions of misfit vectors $\vec{\delta(r)}$ are shown by
filled circles for the core structures shown in Fig. \ref{fign-ddmap} (a) and (b), together with the contour maps of the GSFE.
The white circles in the  \pl case correspond to a segment
of \pz orientation in the non-planar SF in Fig. \ref{fign-ddmap} (a).
The crosses mark the minima of the GSFE.
(b) The GSFE landscape along the straight path for the \pl and \pz cases.
}
\label{fign-misfit2}
\end{figure}

Such a small energy difference between the planar and non-planar core shapes is a striking feature,
and it can be explained in terms of GSFE landscape.
The core structure is characterized by
a series of misfit vectors $\vec{\delta(r)}$, defined as
discontinuity of atomistic displacements at position $r$ on the slip plane \cite{bulatov97}.
They are calculated from the differentiated displacements
whose representative arrows in Fig. \ref{fign-ddmap} intersect with the SF.
In the framework of the semidiscrete Peierls-Nabarro model,
the dislocation energy is expressed in terms of 
SF energy $\gamma_{SF}(\delta(r))$
and elastic energy expressed by a quadratic of strain which in turn is approximated by
a finite difference $\nabla \delta(r) \sim \delta(r_1) - \delta(r_2)$.
If misfit vectors $\vec{\delta(r)}$
are regarded as a sequence of points in the
space of displacement, they
essentially behave like an elastic chain embedded in the GSFE landscape
whose two ends are fixed at $\vec{0}$ and $\vec{b}$ \cite{bulatov97}.
In an idealized SF case, many points aggregate at the GSFE minimum which corresponds
to the partial Burgers vector.
Figure \ref{fign-misfit2} (a) shows distributions of $\vec{\delta(r)}$ 
for the configurations shown in Fig. \ref{fign-ddmap} (a) and (b)
which dissociate on the \pl and \pz planes, respectively,
together with the contour maps of the GSFE.
In both cases, the points almost align in a straight line. 
In the \pl case, the points do not aggregate at the GSFE minimum
and the path only slightly bends towards the minimum,
indicating that the effect of GSFE gradient is weaker than that of the elastic energy \cite{note:straight}.

The fact that the misfit vectors are nearly parallel to the Burgers vector on
both planes implies that the two kinds of SF plane can be joined to
form a non-planar SF with a very
low energy cost, which is proportional to
the square of
discontinuous jump in the misfit vectors at the junction.
In fcc case, the discontinuity is as large as $b/3$
and such non-planar SF is rarely observed, except 
when the SF width is very narrow \cite{al2005}.
Figure \ref{fign-misfit2} (b) shows GSFE landscape along the
straight path for the \pl and \pz cases. 
Around a displacement $\vca/2$, the \pz GSFE is about $0.02$ J/m$^2$ lower than the \pl case,
  while the \pl GSFE is lower at around $0$ and $\vca$.
Correspondingly, all the core structures shown in Fig. \ref{fign-ddmap} have a \pz
segment of SF near their center, and
the core structures (a) and (c) have \pl SF segments on both sides.
These asymmetric, non-planar core shapes come from the crossover of
GSFE on two slip planes and the low energy cost to form SF junctions.
Similar shape of the \pl screw dislocation was also obtained using an
interatomic potential which reproduces the \ab results of pyramidal
GSFE \cite{wu15md}.

\begin{figure}[h]
\includegraphics[width=10cm, bb=0 0 1467 766]{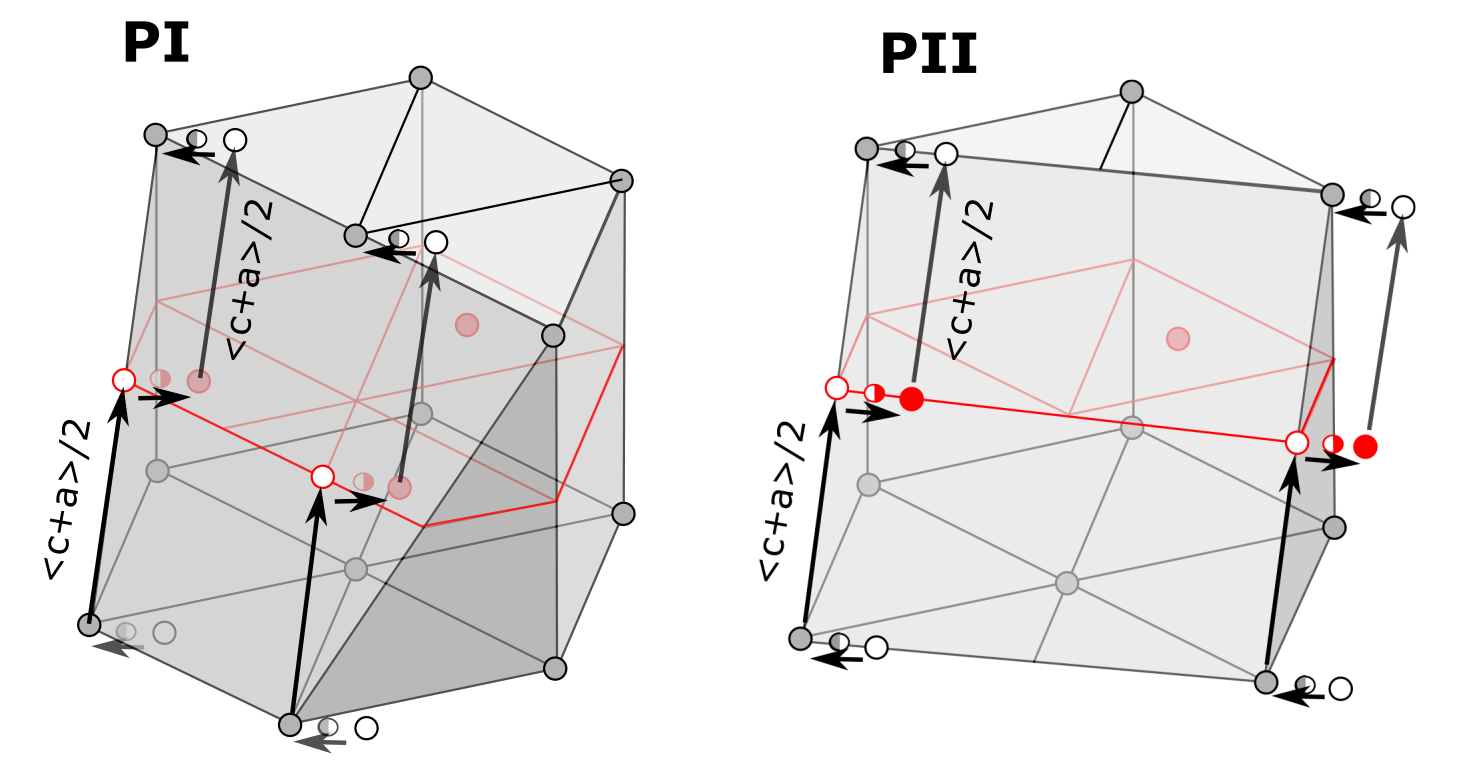}
\caption{
Schematic of atomistic shuffling motion, by which a 
pyramidal SF plane migrates.
Filled and empty spheres denote on-lattice and off-lattice atoms, respectively.
Half-filled spheres are atoms at the intermediate position.
Short arrows show shuffling motion, which induces stacking fault migration.
}
\label{fign-migsf}
\end{figure}

The migration of the SF normal to its plane is unusual, since such
a process usually requires the collective motion of atoms and 
necessitates high activation energy.
We show that a mechanism peculiar to the hexagonal metals enables the migration.
Figure \ref{fign-migsf} shows a schematic of the SF
migration process resulting from the shuffling of atoms.
When atoms on a \pl or \pz plane are shifted by $\vca /2$, they end up
at off-lattice positions.
These atoms can move back to their on-lattice positions by a local shuffling motion.
If all the atoms on the SF execute this shuffling, the SF migrate in its normal direction
by one atomistic layer \cite{tang15}.
In the intermediate configuration, atoms align in a row along the $\vca$ direction with a period of $\vca /2$. 
The energy of this intermediate configuration is found to be almost the same as that of the initial and the final configurations. 
Furthermore, individual shuffling movements of each atom
only moderately change the distance between the atoms,
indicating that shuffling does not require a
collective motion.

\begin{figure}[h]
\includegraphics[width=17cm, bb=0 0 1335 936]{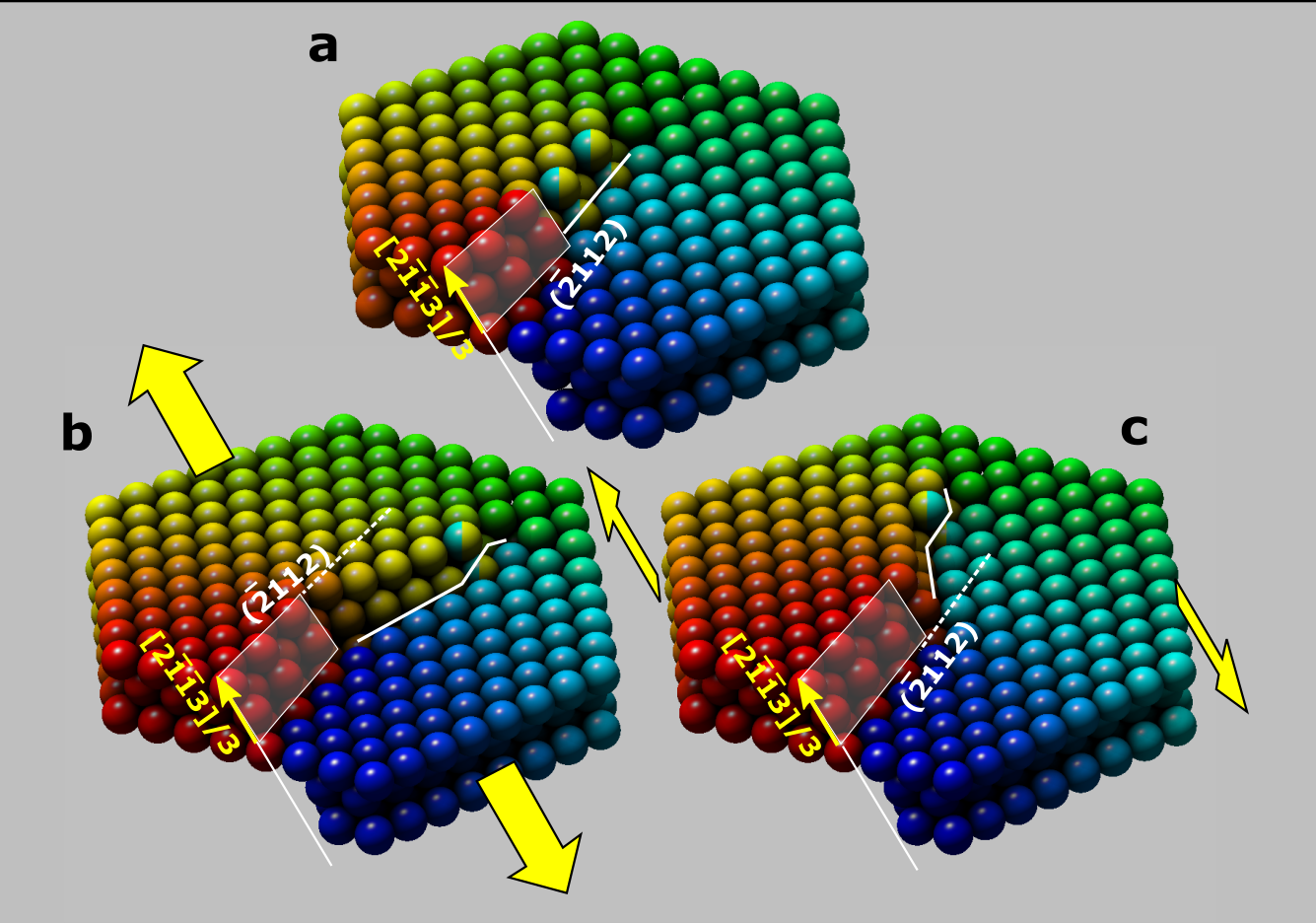}
\caption{
    Atomistic core structures of $\vca$ screw dislocations:
	\figl{a} stable core structure when no strain is applied, and
	after uniform strain in the direction of
	\figl{b} $\langle 2\bar{1}\bar{1}3\rangle (\bar{1}011)$ and
	\figl{c} $\langle 2\bar{1}\bar{1}3\rangle (\bar{1}101)$  is
	applied and relaxed using the \ab method.
	The bold yellow arrows show the direction of the applied shear strain.
	Dashed and solid lines denote the positions of
	the SF before and after the application of strain, respectively.
	The color of atoms corresponds to their displacement along the Burgers vector.
	The atoms shown by two colors are half-shuffled atoms.
}
\label{fign-stress}
\end{figure}

The flexibility of the SF, i.e., the ability to
migrate and assume a non-planar shape, indicates that
the conventional wisdom of the identity between the slip plane
and the SF plane is not applicable in Mg.
The actual plane of slip is determined by 
the anisotropy of the lattice friction and the Schmid factor.
Figure \ref{fign-stress} shows the \ab result for dislocation configurations
when a uniform strain is applied 
to the stable core configuration dissociated on the \pz plane.
A uniform strain in the 
$\vca (01\bar{1}1)$ and $\vca (1\bar{1}01)$ directions
is applied and gradually increased, $0.1\%$ at a time
and up to $1.0\%$ and  \ab relaxations are carried out at each step.
One can clearly see that the SF plane changes from \pz to \pl
according to the applied stress.
No strong preference for the \pz over the \pl slip plane is observed. 

This finding resolves the discrepancy between a recent \ab result
\cite{ghazi-dft} where a \pz screw
dislocation was found to be more stable than a \pl screw dislocation, and
experimental results indicating
that more \pl than \pz slip traces are observed in some cases
\cite{obara73,fan15}.
When a dislocation is at rest
with no applied stress, it dissociates on the \pz plane.
When a stress
is applied, it is more likely that the dislocation glides on the \pl plane, as 
has also been shown by MD simulations \cite{tang14,tang14-2}.
For loading in the $\vc$ direction, two types of \pl slip
$\vca (01\bar{1}1)$ and $\vca (1\bar{1}01)$ have the same Schmid factor
and frequent cross-slips between them is expected, as has been observed in \cite{xie16}.
Transformation of core structure in a dislocation line requires
a nucleation of transformed segment with some critical length.
Assuming that the critical length and the energy barrier for the transformation
are of the order of $10b$ and $2$ meV/\AA, respectively,
the activation energy is about $100$ meV.
At the ambient temperature, the waiting time for this activation is about several pico seconds and core structure will transform in response to the applied stress quickly enough to be 
observed in MD simulations with modest strain rate.

One of the most imminent questions concerning $\vca$ dislocations is the
source mechanism; it is more plausible to
assume that $\vca$ dislocations are generated by a reaction between 
two types of dislocation than directly from  some type of dislocation source.
Several hypothetical mechanisms have 
been proposed, such as a reaction between $\vc$ and $\va$ 
dislocations \cite{source-hcp} and 
the emission of a partial from a vacancy/interstitial loop \cite{agnew-i1}. 
The flexibility of the partial screw dislocation
and the SF shown in the present work will open up
many more possibilities for candidate mechanisms.



\newcommand{\citt}[5]{{#1} #5;#2:#3}
\newcommand{\cit}[5]{ \citt{#1}{#2}{#3}{#4}{#5}.}

\def\prl{Phys. Rev. Lett.}
\def\prb{Phys. Rev. B}
\def\pre{Phys. Rev. E}
\def\pr{Phys. Rev.}
\def\philmag{Philos. Mag.}
\def\actamat{Acta Mater.}
\def\actamet{Acta Metall.}
\def\smat{Scripta Mater.}
\def\jpsj{J. Phys. Soc. Jpn.}
\def\jnm{J. Nucl. Mater.}
\def\ijp{Int.J.Plastcity}
\def\msmse{Model. Simul. Mater. Sci. Eng.}
\def\progms{Prog. Mater. Sci.}
\def\msea{Mater. Sci. Eng. A}

\begin{thebibliography}{99}
\bibitem{note:twin-vector}
{The displacement for the \pl plane is very close to the displacement vector of the $\polll$ extension twin, $\eta \langle 01\bar{1}2\rangle /2$ with $\eta= 3a^2 /(4c^2+3a^2) \sim 0.45$, and the SF can be regarded as two successive twin boundaries.}
\end{thebibliography}
\end{document}